# Chemical abundances in the dwarf galaxy NGC 4163 based on the nebular and auroral emission lines

I. A. Zinchenko*[1,2] | L. S. Pilyugin[2,3]

[1]Faculty of Physics, Ludwig-Maximilians-Universität, Scheinerstr. 1, 81679 Munich, Germany

[2]Main Astronomical Observatory, National Academy of Sciences of Ukraine, 27 Akademika Zabolotnoho St, 03680, Kyiv, Ukraine

[3]Institute of Theoretical Physics and Astronomy, Vilnius University, Sauletekio av. 3, 10257, Vilnius, Lithuania

**Correspondence**
*Igor Zinchenko. Faculty of Physics, Ludwig-Maximilians-Universität, Scheinerstr. 1, 81679 Munich, Germany. Email: Igor.Zinchenko@physik.lmu.de

**Funding Information**
German Research Foundation (DFG), NAS of Ukraine,

We constructed an oxygen abundance map and N/O ratio map of the unusually low excitation dwarf irregular galaxy NGC 4163 based on publicly available spectroscopy obtained by the MaNGA survey. We detected auroral emission line [O II]$\lambda\lambda$7320,7330 which allows us to measure chemical abundance by direct $T_e$ method. We found that the scatter of the oxygen abundance derived by the strong line method is large. The oxygen abundances $12 + \log(O/H)$ derived by strong line method vary from $\sim 7.3$ to $\sim 7.8$ with a mean value of $\sim 7.55$. The oxygen abundances derived in two apertures of 2 arcseconds by the direct $T_e$ method using our measurements of the $O^+$ auroral line is about 7.8 dex. The nitrogen-to-oxygen ratio log(N/O) of about -1.5 is typical value for a low metallicity galaxy, maybe slightly shifted towards higher N/O ratios with respect to the N/O values in the H II regions in nearby galaxies. An unusual negative trend between log(N/O) and oxygen abundance is detected. NGC 4163 is a gas-poor galaxy with a neutral atomic gas mass fraction of around 0.25. The oxygen abundance in the galaxy is only around 0.1 of the oxygen abundance potentially attainable in a galaxy with such a gas mass fraction. The low metallicity coupled with the low gas mass fraction implies that either the metallicity of the interstellar medium of the galaxy was reduced by pristine gas infall in the recent epoch or the evolution of this galaxy was accompanied by strong galactic winds.

## 1 | INTRODUCTION

Measurements of the dwarf irregular galaxy NGC 4163 were carried in optical, infrared, and radio wavelengths. Different characteristics of NGC 4163 were determined and discussed in many works (e.g., Berg et al., 2012; Dalcanton et al., 2009; Huchtmeier & Richter, 1988; Hunter et al., 2011, 2018; Lelli, Verheijen, & Fraternali, 2014; McQuinn, Mitchell, & Skillman, 2015; Moustakas & Kennicutt, 2006; Stilp et al., 2013; Swaters & Balcells, 2002). NGC 4163 is included in a LITTLE THINGS (Local Irregulars That Trace Luminosity Extremes, The H I Nearby Galaxy Survey) sample of galaxies (Hunter et al., 2012). The stellar and the H I masses of NGC 4163 are $M_\star \sim 10^{7.3} M_\odot$ and $M_{HI} = 10^{6.7} M_\odot$ (Stilp et al., 2013). The isophotal radius $R_{25}$ of NGC 4163 in the B band is 57.5 arcsec (Swaters & Balcells, 2002) or 0.8 kpc for a luminosity distance of $2.87 \pm 0.04$ Mpc (Dalcanton et al., 2009). The recent star formation rate in NGC 4163 is low, only around 5% of the stellar population of NGC 4163 was formed in the past $\sim$5 Gyrs, while the average fraction of young stellar population in nearby isolated dwarf galaxies is 25% (Benítez-Llambay et al., 2015).

The chemical abundance of a galaxy provides important information about its evolution. Berg et al. (2012) measured the spectrum of an H II region in NGC 4163 and found the oxygen abundance of the NGC 4163 to be 12 + log(O/H) = 7.56. It is known that high values of the excitation parameter $P^1$ correspond to galaxies of low metallicity (Stasińska, Izotov, Morisset, & Guseva, 2015). The excitation parameter reflects the strength of the high-excitation zone in H II region and, therefore, connected with a number of physical parameters of H II region such as metallicity, hardness of the radiation from ionization source, whether H II region is density-bounded.

However, the excitation parameter of the H II region of NGC 4163 measured by Berg et al. (2012) is very low, $P \sim 0.15$. The integrated spectrum of NGC 4163 obtained by Moustakas & Kennicutt (2006) also shows a low excitation, $P \sim 0.3$. Meanwhile, other low metallicity galaxies (12+log(O/H) < 7.8) have $P > 0.5$ (Stasińska et al., 2015, see also Fig. 5 ). Thus, this low-metallicity galaxy shows an unusually low excitation.

Recently NGC 4163 was observed using integral field unit (IFU) spectroscopy in the framework of the "Mapping Nearby Galaxies at APO" (MaNGA) survey of the Sloan Digital Sky Survey (SDSS-IV) (Albareti et al., 2017). The MaNGA observation identifier of NGC 4163 is 8554-12704. The size of a spaxel at the distance of NGC 4163 is $\sim$ 7 pc. However, the size of individual spaxels (0.5 arcseconds) is significantly smaller compared to a median spatial resolution of MaNGA datacubes, FWHM = 2.54 arcseconds (Law et al., 2016). Thus, the spatial resolution of NGC 4163 datacube is $\sim$ 35 pc. Therefore, the MaNGA observations make it possible to investigate the spatial variations of the radiation in different lines (and, consequently, the variation of the excitation parameter) across this galaxy. The MaNGA spectroscopy allows one also to construct an abundance map for the galaxy and to examine both the presence of a global abundance gradient and of local variations of the abundance.

The irregular galaxies NGC 4163, NGC 4214, and the fainter UGCA 276 (DDO 113) lie potentially within 100 kpc of each other (Hunter et al., 2018). The uncertainty in the distances give an uncertainty in the radial separation of 200 kpc. There are no obvious signs of an interaction in the optical or H I, but the distances between those galaxies are small enough that it is plausible that gravitational effects between them may have occurred at some time in the past (Hunter et al., 2018). It is interesting to note that NGC 4163 was included in a list of dwarf irregular galaxies that are candidates for having formed as tidal dwarfs (Hunter, Hunsberger, & Roye, 2000). The stellar mass of UGCA 276 is $M_\star \sim 10^7$ M$_\odot$ (Weisz et al., 2011) and its H I mass is $M_{\text{HI}} < 4 \times 10^5$ M$_\odot$ (Ott et al., 2012). The stellar and H I masses of NGC 4214 are $M_\star = 10^{8.9}$ M$_\odot$ and $M_{\text{HI}} = 10^{8.5}$ M$_\odot$ (Stilp et al., 2013), Thus, the baryonic mass of NGC 4163 is significantly higher than that of UGCA 276 and significantly lower than the mass of NGC 4214. It should be noted that the gas mass fraction is lowest for the galaxy of the lowest mass, UGCA 276, and largest for the most massive galaxy, NGC 4214.

---
[1]Throughout the paper, we will use the following standard notations for the line intensities:
$R_2 = I_{[\text{O II}]\lambda3727+\lambda3729}/I_{\text{H}\beta}$,
$N_2 = I_{[\text{N II}]\lambda6548+\lambda6584}/I_{\text{H}\beta} = 1.33 I_{[\text{N II}]\lambda6584}/I_{\text{H}\beta}$,
$S_2 = I_{[\text{S II}]\lambda6717+\lambda6731}/I_{\text{H}\beta}$,
$R_3 = I_{[\text{O III}]\lambda4959+\lambda5007}/I_{\text{H}\beta} = 1.33 I_{[\text{O III}]\lambda5007}/I_{\text{H}\beta}$.
Based on these definitions, the excitation parameter $P$ is expressed as $P = R_3/R_{23} = R_3/(R_2 + R_3)$.

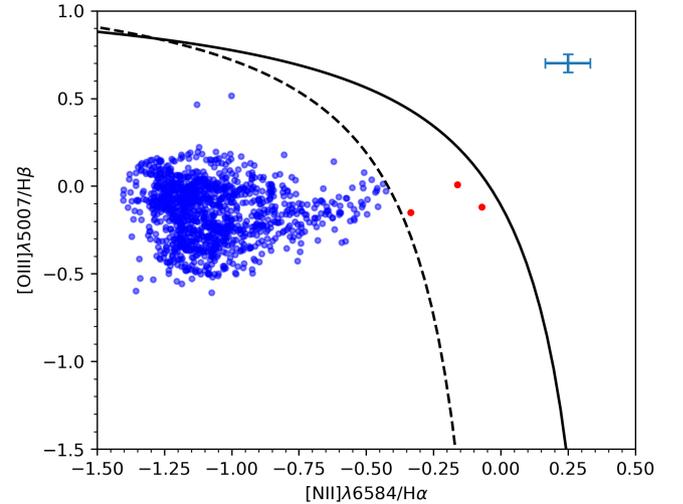

**FIGURE 1** The BPT diagram for NGC 4163. Blue circles show spaxels used in further analysis while red circles mark spectra in a composite area that have have been rejected. The solid line is the theoretical upper limit for SF galaxies presented by Kewley et al. (2001). The dashed line is the demarcation line proposed by Kauffmann et al. (2003) as a lower limit for area of AGNs. Error bars in the top right corner represent median errors for data points on both axes.

In this work we determine the chemical abundance map for the unusual dwarf galaxy NGC 4163 based on MaNGA data. The obtained abundance coupled with the available stellar and gas masses of the galaxy allows us to estimate the loss of chemical elements (efficiency of the galactic wind) in this dwarf galaxy. We compare the abundance and loss of chemical elements of NGC 4163 with that for the potential disturber, NGC 4214.

The paper is organized in the following way. The data for NGC 4163 are described in Section 2. In Section 3 the abundance is determined. The discussion is given in Section 4. Section 5 contains a brief summary.

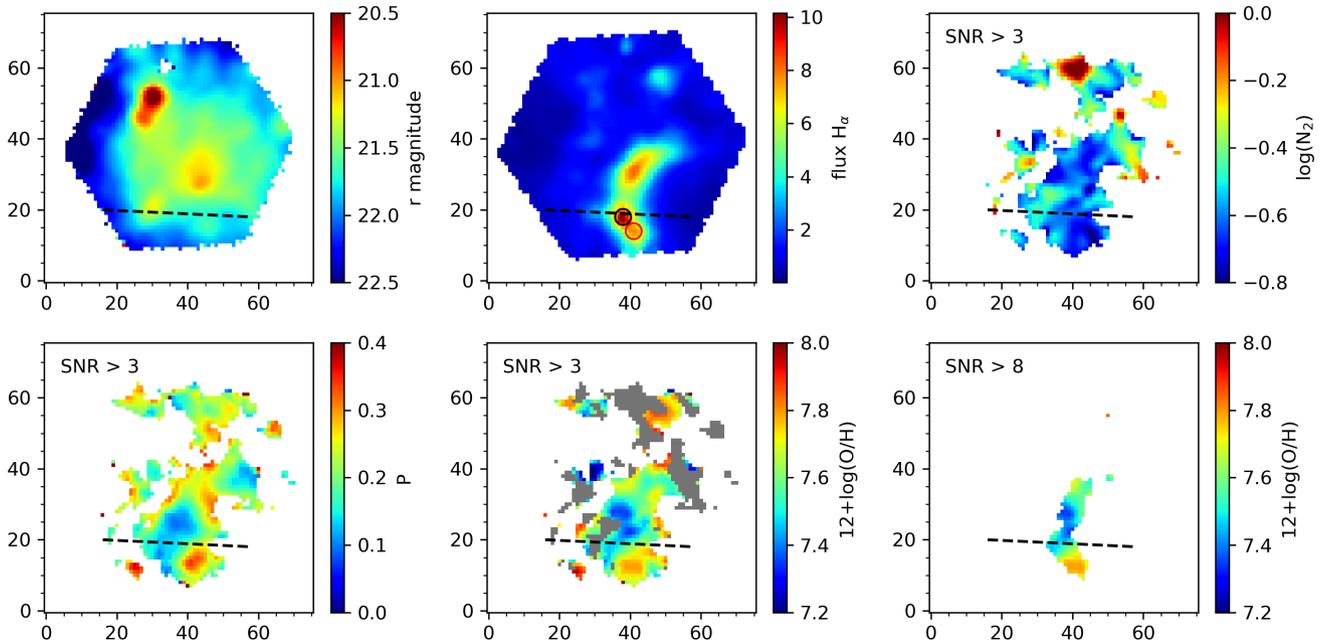

**FIGURE 2** Distribution of the surface brightness (top left panel, the H$\alpha$ flux plotted in units of $10^{-17}$ erg s$^{-1}$ cm$^{-2}$ spaxel$^{-1}$ (top middle panel), the $N_2$ line flux in the logarithmic scale (top right panel), the excitation parameter $P$ (bottom left panel), oxygen abundance for the spaxel spectra with log $N_2 < -0.5$ and SNR > 3 (bottom middle panel), and oxygen abundance for the spaxel spectra with log $N_2 < -0.5$ and SNR > 8 (bottom right panel) across the image of the galaxy NGC 4163 in pixel coordinates. The spaxels with SNR greater than threshold but with log $N_2 > -0.5$ in the oxygen abundance plots are colour-coded in gray. The dashed line is the slit position of the measurement from Berg et al. (2012). Black and red circles represent two apertures with detected [O II]$\lambda$7320,7330 auroral lines in the MaNGA spectra.

## 2 | DATA

The IFU data from the SDSS-IV MaNGA DR15 survey (Albareti et al., 2017) offer the possibility to measure distributions of the observed velocities, surface brightness, and the emission line fluxes across the image of the galaxy. The spectrum of each spaxel from the MaNGA datacube 8554-12704 of NGC 4163 is reduced in the manner described in Zinchenko, Pilyugin, Grebel, Sánchez, & Vílchez (2016); Zinchenko et al. (2021). In brief, the stellar background in all spaxels is fitted using the STARLIGHT code (Asari et al., 2007; Cid Fernandes, Mateus, Sodré, Stasińska, & Gomes, 2005; Mateus et al., 2006). To fit the stellar spectra, we used simple stellar population (SSP) spectra from the evolutionary synthesis models by Bruzual & Charlot (2003) with ages from 1 Myr up to 13 Gyr and metallicities $Z$ of 0.004, 0.02, and 0.05. The resulting stellar spectrum is subtracted from the observed spectrum to obtain a pure gas spectrum. The line intensities in the gas spectrum were measured by fitting single Gaussian profiles on the pure emission spectra by our code ELF3D.

For each spaxel of 8554-12704 datacube we have measured the fluxes of the [O II]$\lambda$3727+$\lambda$3729, H$\beta$, [O III]$\lambda$4959, [O III]$\lambda$5007, [N II]$\lambda$6548, H$\alpha$, [N II]$\lambda$6584, [S II]$\lambda$6717, and [S II]$\lambda$6731 lines and found 1174 spaxels with signal-to-noise ratio (SNR) higher than 3 for each of those lines (and 226 spaxels with a SNR > 8).

We corrected the emission line fluxes for interstellar reddening using the analytical approximation of the interstellar reddening curve derived by Izotov, Thuan, & Lipovetsky (1994) and the theoretical H$\alpha$/H$\beta$ ratio equals to 2.87. Although this approximation is based on Whitford (1958) curve, the effect of applying more recent reddening laws on the derived chemical abundance is small in the optical range (see, e.g., Izotov, Thuan, & Stasińska, 2007). The H$\alpha$/H$\beta$ ratio is obtained for the recombination case B for an electron temperature of 10 000 K at the limit of the low density (Osterbrock & Ferland, 2006).

We use the standard diagnostic diagram of the [N II]$\lambda$6584/H$\alpha$ versus the [O III]$\lambda$5007/H$\beta$ line ratios suggested by Baldwin, Phillips, & Terlevich (1981), which is known as the BPT classification diagram, to separate H II region-like objects and AGN-like objects. We adopted the demarcation line of Kauffmann et al. (2003) between H II regions and AGNs. We found no evidence of the presence of AGN-like objects in the studied area. Only 3 spaxels with AGN-like line ratios have been detected, all in the composite area between demarcation lines of Kauffmann et al. (2003) and Kewley et al. (2001) and with S/N ratio between 3 and 4 (Fig. 1 ).

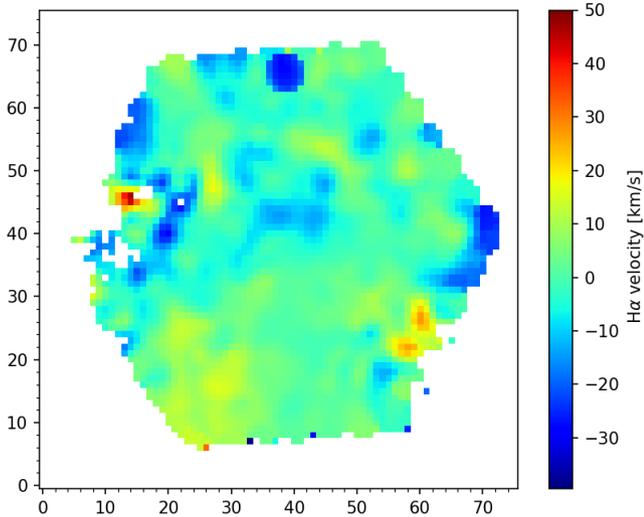

**FIGURE 3** The line of sight velocity map of the gas in NGC 4163 traced by $H_\alpha$ emission line.

The surface brightness in the SDSS *g* and *r* bands for each spaxel was obtained from broadband SDSS images created from the data cube. Top left panel of Fig. 2  shows the distribution of the measured surface brightness across the image of the galaxy NGC 4163 in pixel coordinates. The value of the surface brightness is colour-coded.

The position of the photometric centre of the galaxy was taken from the NASA Extragalactic Database (NED)[2] (R.A. = 183.038150°, Dec = 36.169196° in J2000 coordinates). The position angle of the major axis PA = 15° was taken from Hunter et al. (2011). In principle, the position of the photometric centre of the galaxy, its inclination angle *i*, and the position angle of the major photometric axis can be obtained from the analysis of the measured surface brightness map. Still, we do not determine those values for NGC 4163; instead, we take those values from the literature because of the following reason. The optical radius of the NGC 4163 is 57.5 arcsec (Swaters & Balcells, 2002). Our deprojected photometric map extends to less than ∼ 25 arcsec (less than ∼ 45 MaNGA spaxels). This prevents us from obtaining an accurate determination of the position of the photometric centre of the galaxy by analyzing IFU photometric map. It should be noted that the position of the photometric centre of NGC 4163 is not well defined. Its coordinates reported in different works

(Hunter et al., 2011; Karachentsev, Karachentseva, Huchtmeier, & Makarov, 2004; McQuinn et al., 2015; Swaters & Balcells, 2002) can differ by several arcseconds.

Top middle panel of Fig. 2  shows the distribution of the measured and dereddened H$\alpha$ fluxes in units of $10^{-17}$ erg/s/cm$^2$/spaxel across the image of the galaxy NGC 4163. Top right panel of Fig. 2  shows the distribution of the measured and dereddened $N_2$ line intensity in the logarithmic scale. Bottom left panel of Fig. 2  shows the distribution of the excitation parameter *P* across the image of the galaxy NGC 4163.

Bottom middle panel of Fig. 2  shows the oxygen abundance map for spaxel spectra with log $N_2 < -0.5$ and S/N > 3 in all lines needed for oxygen abundance determination. The detailed description of the oxygen abundance determination is given below. Bottom left panel shows the same plot as in the previous panel but only for spaxels with SNR > 8. One can notice that changes of *P* show some correlation with variations of the oxygen abundance while in general those properties of H II regions anti-correlate. However, this anti-correlation is not tight and at a given value of oxygen abundance a significant scatter in *P* can be observed (see, e.g., Pilyugin & Grebel (2016)).

The measured wavelength of the emission lines provides the gas velocity in each region (spaxel). The emission line profile is fitted by a Gaussian, and the position of the centre of the Gaussian is adopted as the wavelength of the emission line. As can be seen in Fig. 3 , the variation in the velocities obtained from the H$\alpha$ emission line measurements are small and the measured velocity field of the NGC 4163 is not regular enough, which prevents us from determining the position of the kinematic centre of the galaxy, its inclination angle, the position angle of the the major kinematic axis, and the rotation curve.

## 3 | ABUNDANCES IN THE GALAXY NGC 4163

### 3.1 | Properties of the emission lines in the spaxel spectra of NGC 4163

First, we compare the properties of the emission lines in the spaxel spectra of NGC 4163 with those in other MaNGA galaxies and in H II regions in nearby spiral and irregular galaxies. For the comparison with other MaNGA galaxies we selected a sample of the MaNGA DR15 star-forming galaxies. According to the commonly used BPT diagnostic diagram (Baldwin et al., 1981), we selected 1851155 spectra with SNR > 3 in all of the [O II]$\lambda$3727+$\lambda$3729, H$\beta$, [O III]$\lambda$4959, [O III]$\lambda$5007, [N II]$\lambda$6548, H$\alpha$, [N II]$\lambda$6584, [S II]$\lambda$6717, and [S II]$\lambda$6731 lines from 3483 galaxies, which have H II-region-like spectra. It should be noted that we do not disentangle irregular and spiral galaxies galaxies, although, the majority of galaxies in our sample are spirals. We also make comparison between H II regions in NGC 4163 and the H II regions in nearby galaxies.

---
[2]The NASA/IPAC Extragalactic Database (NED) is operated by the Jet Propulsion Laboratory, California Institute of Technology, under contract with the National Aeronautics and Space Administration. http://ned.ipac.caltech.edu/

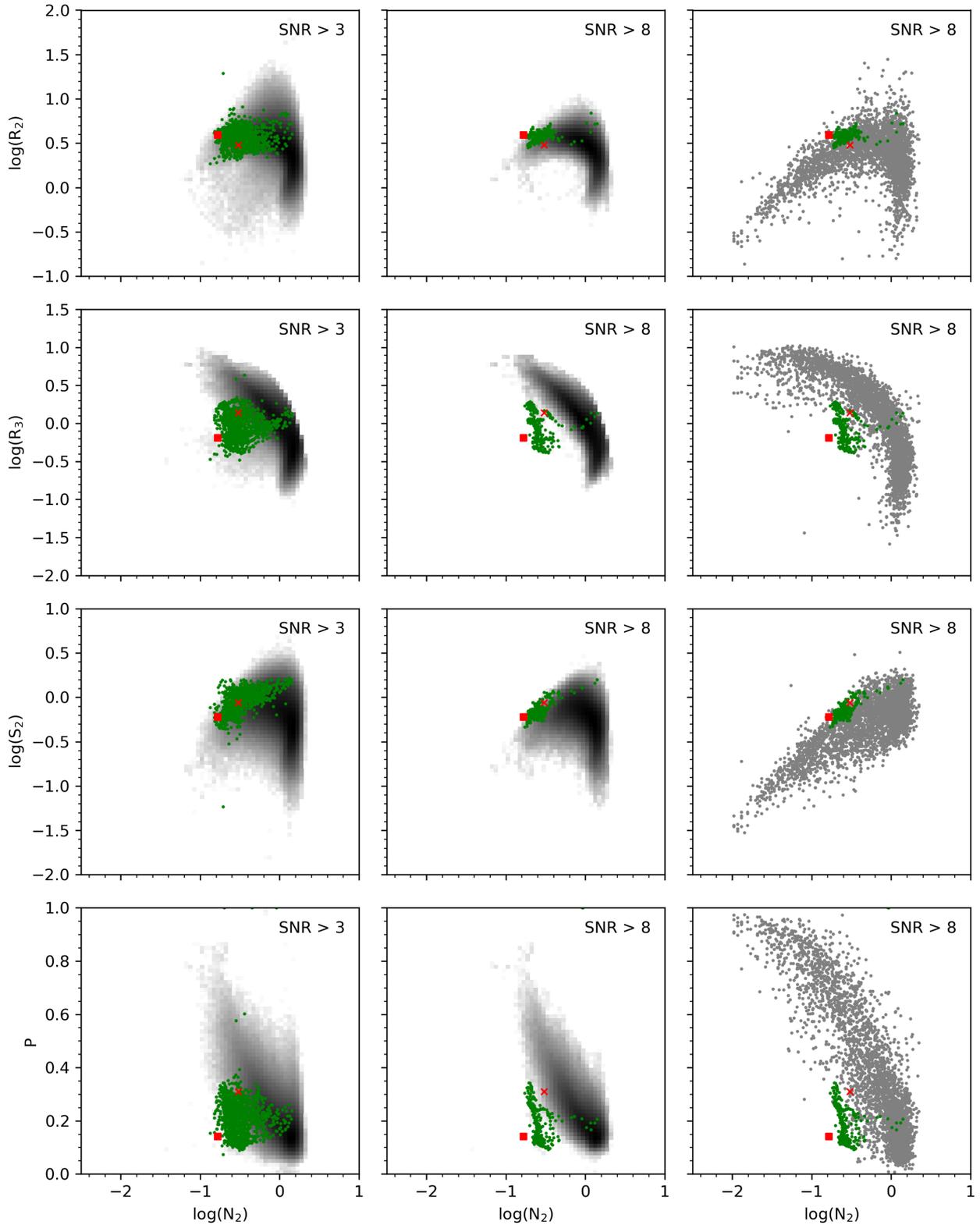

**FIGURE 4** Left and middle panels show the comparison of the emission line fluxes in the spaxel spectra of NGC 4163 (green points) with those in a large sample of other MaNGA galaxies shown by the 2D histogram colour-coded in gray. The left panels show spectra with SNR > 3 while the middle panels present spectra with SNR > 8. The right panels show the comparison of the emission line fluxes in MaNGA spaxel spectra with SNR > 8 of NGC 4163 with the fluxes in the slit spectra of H II regions in nearby galaxies (gray points). The red square denotes the slit spectrum of the region in NGC 4163 from Berg et al. (2012), the red cross marks the locus of the integrated spectrum of NGC 4163 from Moustakas & Kennicutt (2006).

In the top left panel of Fig. 4 we show the $N_2 - R_2$ diagram for the spectra of individual spaxels in NGC 4163 with SNR > 3 which are shown by the green points. As a comparison, we plot a distribution of all spaxels in our sample of of MaNGA DR15 galaxies which classified as H II-region-like spectra with SNR > 3. The red square denotes the slit spectrum of the region in NGC 4163 measured by Berg et al. (2012); the red cross marks the integrated spectrum of NGC 4163 from Moustakas & Kennicutt (2006).

The other rows in Fig. 4 show the $N_2 - R_3$, the $N_2 - S_2$, and the $N_2 - P$ diagrams, constructed in a similar way. The middle panels of Fig. 4 show the same as the left panels, but for spectra with SNR > 8.

In our previous studies, we compiled slit spectra of H II regions in nearby spiral and irregular galaxies (Pilyugin & Grebel, 2016; Pilyugin, Grebel, & Kniazev, 2014; Pilyugin, Grebel, & Mattsson, 2012). This sample includes 3440 spectra of individual H II regions in irregular and disk galaxies with measured $R_2$, $R_3$, $N_2$, and $S_2$ lines. The right panels show the comparison of the emission line intensities in the MaNGA spaxel spectra of NGC 4163 with a SNR > 8 with the line intensities in the slit spectra of H II regions in those nearby galaxies (gray points).

Examination of Fig. 4 shows that NGC 4163 has a region with a prominent feature in the gas spectra. Namely, the $R_3$ line in this galaxy is significantly weaker compared to both the spectra of the vast majority of the other MaNGA galaxies and the spectra of H II regions in our sample of nearby galaxies with a similar intensity of the $N_2$ line. This fact implies that the excitation of the spaxel spectra of NGC 4163 is low.

## 3.2 | Abundance determination using direct $T_e$ method

To calculate the chemical abundance by direct $T_e$ method measurement of at least one auroral line is needed. Detection of the auroral line [O III]$\lambda$4363 has been reported in Berg et al. (2012) in NGC 4163 while the sulfur auroral line [S III]$\lambda$6312 was reported to be below detection level. Using auroral [O III]$\lambda$4363 line, Berg et al. (2012) estimated low $T_e$-based oxygen abundance of 12+log(O/H) = 7.56±0.14.

In this work, we carried out searching for [O III]$\lambda$4363, [N II]$\lambda$5755, and [O II]$\lambda\lambda$7320,7330 auroral lines. For detection of auroral line we required SNR > 4 in detecting auroral line as well as in all strong emission lines ([O II]$\lambda$3727+$\lambda$3729, H$\beta$, [O III]$\lambda$4959, [O III]$\lambda$5007, [N II]$\lambda$6548, H$\alpha$, [N II]$\lambda$6584, [S II]$\lambda$6717, and [S II]$\lambda$6731). Using this technique we detected [O II]$\lambda\lambda$7320,7330 emission in the bright H$\alpha$ blob observed by Berg et al. (2012). To measure chemical composition, we divided this blob into two apertures with a diameter of 5 spaxels or 2.5", which roughly corresponds to the spatial resolution of MaNGA data. Those apertures are shown by black (1st aperture) and red (2nd aperture) circles in Fig. 2. Dereddened values of emission line fluxes and corresponding 1-$\sigma$ errors, normalized to H$\beta$, absolute flux in H$\beta$, the extinction coefficient C$_{H\beta}$ are reported in Table 1.

**TABLE 1** Emission line intensities for apertures with detected auroral lines.

| Ion | Aperture 1 | Aperture 2 |
| --- | --- | --- |
| [O II]$\lambda\lambda$3727, 3729 | 3.960± 0.113 | 3.874± 0.118 |
| [Ne III]$\lambda$3889 | — | 0.067± 0.017 |
| H$\gamma$ | 0.456± 0.013 | 0.479± 0.015 |
| [O III]$\lambda$4363 | — | — |
| H$\beta$ | 1.000± 0.011 | 1.000± 0.015 |
| [O III]$\lambda$4959 | 0.185± 0.010 | 0.452± 0.014 |
| [O III]$\lambda$5007 | 0.595± 0.011 | 1.302± 0.016 |
| [N II]$\lambda$5755 | — | — |
| [O I]$\lambda$6300 | 0.078± 0.021 | — |
| [S III]$\lambda$6312 | — | — |
| [N II]$\lambda$6548 | 0.067± 0.012 | 0.072± 0.014 |
| H$\alpha$ | 2.875± 0.019 | 2.876± 0.019 |
| [N II]$\lambda$6584 | 0.181± 0.012 | 0.151± 0.011 |
| [S II]$\lambda$6717 | 0.444± 0.010 | 0.367± 0.010 |
| [S II]$\lambda$6731 | 0.294± 0.010 | 0.251± 0.010 |
| [O II]$\lambda$7320 | 0.050± 0.007 | 0.067± 0.013 |
| [O II]$\lambda$7330 | 0.046± 0.007 | 0.036± 0.010 |
| [S III]$\lambda$9068 | 0.126± 0.009 | 0.132± 0.009 |
| [S III]$\lambda$9530 | 0.333± 0.015 | 0.364± 0.016 |
| Absolute flux H$\beta$† | 69.201± 0.775 | 56.314± 0.869 |
| C$_{H\beta}$ | 0.067± 0.017 | 0.0486± 0.023 |

†In units of $10^{-17}$ erg s$^{-1}$ cm$^{-2}$.

**TABLE 2** Gas properties in apertures with detected auroral lines.

| | Aperture 1 | Aperture 2 |
| --- | --- | --- |
| 12+log(O/H) | 7.83± 0.08 | 7.79± 0.10 |
| log(N/O) | -1.47± 0.04 | -1.51± 0.03 |
| $t_2$ [$10^4$ K] | 1.24± 0.08 | 1.32± 0.13 |
| $n_e$ [cm$^{-3}$] | 50± 14 | 63± 21 |

The electron density was calculated from the [S II]$\lambda$6717/[S II]$\lambda$6731) line ratio using relation from Osterbrock & Ferland (2006).

For the calculation of the chemical abundance, we used a two-zone model of the HII regions, where $t_2$ and $t_3$ are the electron temperatures (in units of $10^4$ K) in the low and high ionization zones, respectively. The electron temperature $t_2$ was calculated directly as a function of the temperature sensitive nebular to auroral line ratio of $O^+$ ion and electron density:

$$t_2 = \frac{a_n}{\log Q_{2,O} + b_n + c_n \log t_2 + d_n t_2}, \quad (1)$$

where $Q_{2,O}$ = [O II]$\lambda\lambda$3727, 3729/[O II]$\lambda\lambda$7320, 7330 is the ratio of nebular to auroral oxygen $O^+$ line intensities and $a_n$, $b_n$, $c_n$, $d_n$ are coefficients which depend on the electron density $n_e$:

$a_n = 8.91697 \times 10^{-1} - 9.06267 \times 10^{-5} n_e + 4.86622 \times 10^{-8} n_e^2$,
$b_n = -9.30335 \times 10^{-1} + 3.98683 \times 10^{-4} n_e - 1.14528 \times 10^{-7} n_e^2$,
$c_n = -1.81420 \times 10^{-1} - 1.05860 \times 10^{-4} n_e + 4.29814 \times 10^{-8} n_e^2$,
$d_n = 2.09975 \times 10^{-2} - 6.797004 \times 10^{-6} n_e + 2.44840 \times 10^{-9} n_e^2$.

This expression is derived from the fitting of the solution for five-level model for the $O^+$ ion using the emission-line analysis package PyNEB v1.1.13 (Luridiana, Morisset, & Shaw, 2015). The fitting is valid in the $t_2$ range from 0.4 to 2.5 and $n_e$ from 1 to 1000 cm$^{-3}$ with accuracy better than 1%. The electron temperature $t_3$ is obtained from the following relation between $t_2$ and $t_3$ (Campbell, Terlevich, & Melnick, 1986; Garnett, 1992):

$$t_2 = 0.7 t_3 + 0.3. \quad (2)$$

Then, the oxygen abundance and N/O ratio were calculated using relations from Pilyugin et al. (2012). The oxygen abundance, N/O ratio, $t_2$, and $n_e$ with corresponding errors are presented in Table 2 for both apertures. Our estimation of the oxygen abundance 12+log(O/H) $\sim$ 7.8 is higher compared to the abundance found by Berg et al. (2012). However, the difference lies within a 2-$\sigma$ confidence interval.

## 3.3 | Abundance determination by calibration

The unusual properties of the emission lines in the spaxel spectra of NGC 4163 noted above should be taken into account when applying calibrations for the abundance determinations. It has been shown that the three-dimensional $R$ calibration from Pilyugin & Grebel (2016) produce reliable abundances (Lara-López et al., 2021; Pilyugin et al., 2018). Distinct relations of the $R$ calibration for the abundance determinations in high- and low-metallicity objects (upper and lower branches) are used. A value of $\log N_2 = -0.6$ is adopted as the criterion separating the ranges of applicability of those relations in the sense that the abundance in an object with $\log N_2 > -0.6$ is estimated through the upper branch calibration relation and the abundance in an object with $\log N_2 < -0.6$ is estimated through the lower branch calibration relation. The ranges of the applicability of the low- and high-metallicity calibration relations overlap, e.g., the lower branch relation is applicable up to $\log N_2 \sim -0.45$. Therefore the error in the abundance due to the misclassification of objects with $N_2$ close to the transition value is not large.

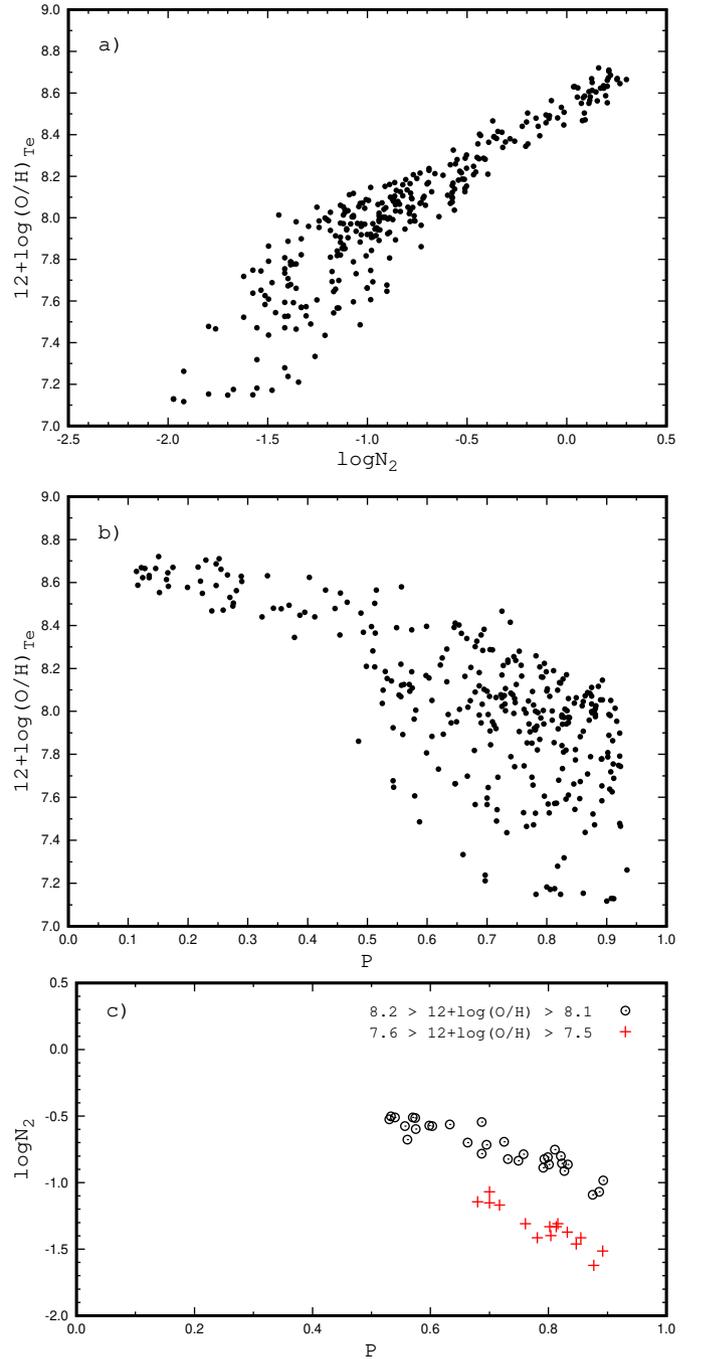

**FIGURE 5** Panel *a* shows the oxygen abundance as a function of the nitrogen emission line $N_2$ flux for a sample of H II regions where the abundance was determined through the direct $T_e$ method with high precision (the calibrating data points). Panel *b* shows the oxygen abundance as a function of the excitation parameter $P$ for that sample of objects. Panel *c* shows the nitrogen emission line $N_2$ flux as a function of the value of the excitation parameter $P$ for H II regions with oxygen abundances from 12+log(O/H) = 8.1 to 8.2 (circles) and for H II regions with 12+log(O/H) from 7.5 to 7.6 (plus signs).

Panel *a* of Fig. 5 shows the oxygen abundance as a function of the nitrogen $N_2$ line intensity for the calibrating datapoints used for creation of the $R$ calibration. Panel *b* shows the oxygen abundance as a function of the excitation parameter $P$ for those objects. The mean value of the oxygen abundance in the calibrating objects with $\log N_2 \sim -0.6$ is $12+\log(O/H) \sim 8.2$. One can see that each calibrating object with metallicity less than $12+\log(O/H) \sim 8.2$ has the value of $\log N_2$ lower than $-0.5$. Similarly, each calibrating object with metallicity less than $12+\log(O/H) \sim 8.2$ has the value of the excitation parameter higher than $\sim 0.5$. Thus, the spectra of the low-metallicity objects are highly excited with low nitrogen line $N_2$ intensity.

Panel *c* of Fig. 5 shows the $N_2$ intensity as a function of the value of the excitation parameter $P$ for calibration objects with oxygen abundances from $12+\log(O/H) = 8.1$ to $8.2$ (circles) and for H II regions with abundances from $12+\log(O/H) = 7.5$ to $7.6$ (plus signs). For a fixed metallicity of the objects, the nitrogen line $N_2$ intensity increases with decreasing excitation parameter.

NGC 4163 has at least one low-metallicity star forming region according to direct metallicity measurement by Berg et al. (2012). However, the nitrogen line $N_2$ in some MaNGA spaxel spectra of this galaxy is rather strong, see Fig. 4. The use of the standard criterion, $\log N_2 = -0.6$, which separates objects on the upper and lower branches would result in the classification of a fraction of the regions in NGC 4163 as a high-metallicity objects since they show $\log N_2 \gtrsim -0.6$. Since the nitrogen line $N_2$ in the MaNGA spaxel spectra of this galaxy may be strong because of the unusually low excitation but not due to a high metallicity, the application of the upper branch calibration for the estimation of oxygen abundances in NGC 4163 is not justified. On the other hand, the lower branch of the $R$ calibration can provide fairly accurate oxygen abundances for objects with $\log N_2$ higher than the adopted limit, up to $\log N_2 \sim -0.5$ (Pilyugin & Grebel, 2016). Thus, in the special case of NGC 4163, the oxygen abundances will be determined only for spaxel spectra with $\log N_2 < -0.5$ where the low branch calibration is workable. Spaxel spectra with $\log N_2 > -0.5$ are rejected.

## 3.4 | Spatial variations of the oxygen abundance and N/O ratio

The normalized histograms of the calibration-based oxygen abundances obtained for individual MaNGA spaxel spectra in NGC 4163 are shown in Fig. 6. It shows that the mean value of the oxygen abundances in NGC 4163 is low, $12 + \log(O/H)_{\mathrm{mean}} \sim 7.57$ if considering all spaxels with SNR > 3 and $\log N_2 < -0.5$. The scatter in the oxygen abundances in NGC 4163 is large, ranging from $12 + \log(O/H) \sim 7.3$ to $\sim 7.8$ with a standard deviation of 0.15 dex. It should be noted that some fraction of spaxels with $\log N_2 > -0.5$ could have larger oxygen abundance values which could lean to even larger spatial variation of the oxygen abundance.

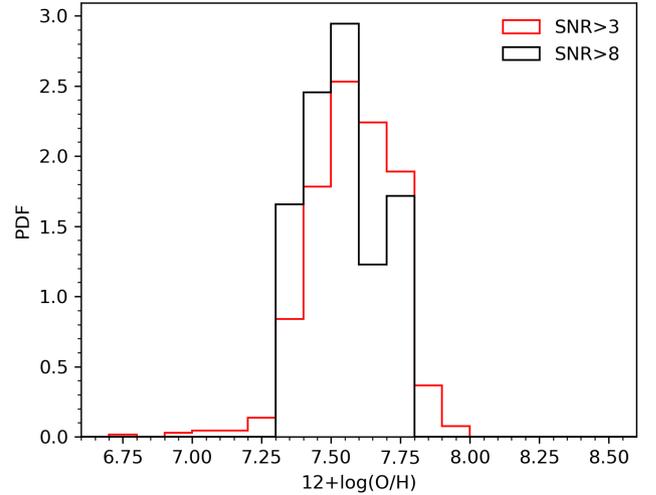

**FIGURE 6** Normalized histograms of oxygen abundances based on MaNGA spectra of individual spaxels in NGC 4163. The red line shows the distribution of abundances based on spaxel spectra with $\log N_2 < -0.5$ and SNR > 3. The black line stands for spaxel spectra with SNR > 8.

To verify the obtained properties of the oxygen abundance distribution derived by the $R$ calibration in NGC 4163, we consider oxygen abundances based on spaxel spectra with higher SNR of > 8. The mean value of the oxygen abundances in NGC 4163 in this case is $12 + \log(O/H)_{\mathrm{mean}} \sim 7.53$, i.e., lower by $\sim 0.04$ dex in comparison to the mean value for the full set of abundances considered before. The scatter in the oxygen abundances does not change significantly and remains large, with standard deviation equal to 0.13 dex. This suggests that the obtained scatter cannot be solely attributed to uncertainties in the line flux measurements.

In Fig. 7 we show the O/H – N/O diagram for the abundances calculated for MaNGA spaxel spectra with $\log N_2 < -0.5$. In this case, we used a version of $R$ calibration for the determination of the N/O ratio from Pilyugin & Grebel (2016). The spatial distribution of the N/O ratio is shown in Fig. 8. Green points correspond to the spaxels with SNR > 3 while red points represent the spaxels with SNR > 8. The grey points denote the abundances in a sample of H II regions in nearby galaxies estimated through the $R$ calibration with ordinary separation criterion for lower and upper branches of $\log N_2 = -0.6$. The sample is taken from the compilation of spectra from Pilyugin et al. (2014, 2012) and Pilyugin & Grebel (2016). The nitrogen-to-oxygen ratio in NGC 4163 is slightly shifted towards higher N/O ratios with respect to the general trend in the O/H – N/O diagram outlined by H II regions in nearby galaxies. However, both oxygen abundance and N/O ratio are close to the ones determined by Berg et al. (2012) from a slit spectrum via the direct $T_e$ method which is shown by the black circle in Fig. 7 .

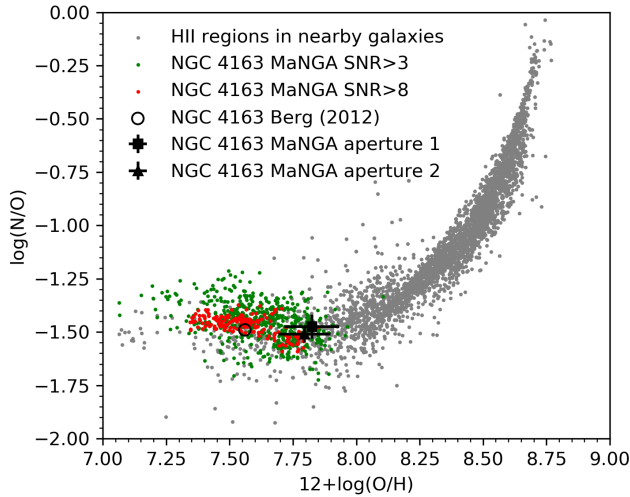
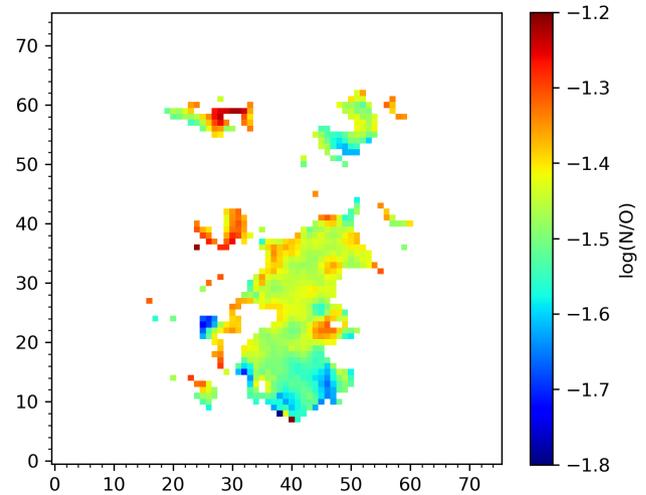

**FIGURE 7** O/H – N/O diagram for the abundances calculated for MaNGA spaxel spectra with $\log N_2 < -0.5$. Green points corresponds to the spaxels with SNR > 3, red points corresponds to the spaxels with SNR > 8. For the comparison, the abundances in H II regions in nearby galaxies are shown by gray points. The black circle denotes the abundance in the region in NGC 4163 determined by Berg et al. (2012) from a slit spectrum through the direct $T_e$ method. Black square and triangle represent abundances calculated by the $T_e$ method in this work.

**FIGURE 8** The spatial distribution of N/O ratio calculated using $R$ calibration for MaNGA spaxel spectra with $\log N_2 < -0.5$ and SNR > 3.

Thus, the mean value of the oxygen abundances in NGC 4163 obtained from the MaNGA spaxel spectra through the $R$ calibration is low and agrees well with the value obtained through the direct method for a single region in that galaxy by Berg et al. (2012). Importantly, while the spatial fluctuations of the oxygen abundance are large, there is no appreciable radial or azimuthal oxygen abundance gradient across NGC 4163.

It is interesting to note that the oxygen abundance in two apertures with detected auroral $O^+$ line is slightly higher compared to the estimation by $O^{++}$ auroral line carried out by Berg et al. (2012). However, this discrepancy lies within 2-$\sigma$ confidence interval and also within the range of estimations by the $R$ calibration in the individual spaxels. Meanwhile, N/O ratios calculated using $R$ calibration and direct $T_e$ method do not show significant difference.

Another interesting feature of the O/H – N/O diagram for NGC 4163 is a negative N/O trend for low oxygen abundance, similar to reported by Kumari, James, Irwin, Amorín, & Pérez-Montero (2018) for NGC 4670 and some other dwarf galaxies. This slope is statistically significant for both samples of spaxels with SNR > 3 and SNR > 8, -0.196±0.023 and -0.224±0.021, respectively. Meanwhile, for a reference sample of H II regions in nearby galaxies with oxygen abundance less than 7.9 dex, the slope is much smaller, -0.074±0.031, with p-value of 0.019. Therefore, the negative slope in case of NGC 4163 is much more statistically significant compared to the reference sample. Detection of a significant negative N/O slope in the O/H – N/O diagram is interesting since, according to current nucleosynthesis models and observations, nitrogen is mainly a primary element at the low metallicity regime (see, e.g., Izotov & Thuan, 1999). And even if nitrogen is a secondary element, a positive trend is expected. On the other hand, a mild negative N/O slope in the O/H – N/O diagram for extremely low-metallicity HII regions has been reported before in Guseva et al. (2011) for the chemical abundances derived with a direct method. Authors suggested that this could be caused by enhanced production of primary nitrogen by rapidly rotating stars at low metallicity.

As has been discussed in Kumari et al. (2018), several mechanisms can be responsible for a negative trend, including a localized nitrogen enrichment of the interstellar medium from the ejecta of WR stars, an inflow of the low-metallicity gas, and varying star formation efficiency (SFE). Analyzing NGC 4163 datacube, we did not find any significant traces of WR stars or gas outflow with high velocity. However, the low spectral resolution of MaNGA spectra makes a very difficult detailed analysis of gas outflow from the bright H II regions. Thus, further observations with higher spectral resolution and signal-to-noise ratio may be needed for quantifying the role of gas outflow and WR stars in this galaxy.

For NGC 4670, Kumari et al. (2018) reported that the spaxels with lower metallicity and higher N/O show higher H$\alpha$ fluxes which suggest a connection of the N/O enhancement with the very recent enhancement of the star formation rate (SFR). However, we did not find the such correlation for NGC 4163 (see the distribution of the spaxels with different SNR in Fig. 7 as an example). This implies that the negative trend in the O/H – N/O diagram cannot be caused only by the recent enhancement of the SFR.

On the other hand, numerical simulations of Köppen & Hensler (2005) predict that inflow of a metal-poor gas can cause significant decreasing of the oxygen abundance but much smaller changes in N/O ratio. Moreover, Köppen & Hensler (2005) models predict that a part of the time track in the O/H – N/O diagram after the accretion event may have a negative slope.

Thus, the negative trend in the O/H – N/O diagram for NGC 4163 may be caused by the inflow of metal-poor gas as discussed in Köppen & Hensler (2005) and Amorín, Pérez-Montero, & Vílchez (2010) or by the different star formation efficiency in different parts of a galaxy as suggested by Kumari et al. (2018) based on the chemical models of Vincenzo, Belfiore, Maiolino, Matteucci, & Ventura (2016). Both mechanisms can also be responsible for significant spatial variations of the oxygen abundance and N/O ratio.

## 4 | DISCUSSION

The stellar and gas masses of NGC 4163 are low, $M_\star = 10^{7.3} M_\odot$ and $M_{HI} = 10^{6.7} M_\odot$ (Stilp et al., 2013), or $M_\star = 10^8 M_\odot$ and $M_{HI} = 1.5 \times 10^7 M_\odot$ (Lelli et al., 2014). However, the astration level $s$

$$s = \frac{M_\star}{M_\star + M_{gas}} \quad (3)$$

is high or alternatively the gas mass fraction $\mu$

$$\mu = 1 - s = \frac{M_{gas}}{M_\star + M_{gas}} \quad (4)$$

is low. The gas mass fraction in NGC 4163 is $\mu \sim 0.25$ for the values of $M_\star$ and $M_{HI}$ from Stilp et al. (2013) and $\mu \sim 0.17$ for the data from Lelli et al. (2014) taking into account the presence of helium, i.e., $M_{gas} = 1.36 M_{HI}$.

The oxygen abundances determined through the $R$ calibration are compatible to the metallicity scale of H II regions defined by H II regions with abundances obtained through the direct $T_e$ method. The value of the oxygen yield $Y_O \sim 0.0030$ corresponds to the metallicity scale defined by the $T_e$-based abundances (Bresolin, Garnett, & Kennicutt, 2004; Pilyugin, Thuan, & Vílchez, 2007; Pilyugin, Vílchez, & Contini, 2004). A simple model for the chemical evolution of galaxies (a closed-box model with instantaneous recycling, abbreviated "cbm") predicts an oxygen abundance of $12+\log(O/H)_{cbm} = 8.53$ for $\mu = 0.25$. The measured oxygen abundance in NGC 4163, $12+\log(O/H) \sim 7.55$, is only $\sim 0.1$ of the potentially attainable abundance for the astration level in the NGC 4163. This implies that either the evolution of this dwarf galaxy was accompanied by strong galactic winds (where the galaxy would have kept only around 0.1 of the potentially produced oxygen) or the abundance of the interstellar medium was reduced by rather pristine gas infall in the recent epoch. The use of the stellar and gas masses of NGC 4163 from Lelli et al. (2014) results in a similar conclusion, i.e., the galaxy kept only around 0.08 of the potentially produced oxygen. It should be noted that the NGC 4163 lies below metallicity – luminosity and metallicity – mass relations for dwarf galaxies by $\sim 0.2$ dex (Berg et al., 2012). However, such deviation is comparable to the scatters in those relations.

It was noted above that the NGC 4163 has two neighboring galaxies, NGC 4214 and UGCA 276. The baryonic mass of NGC 4163 is significantly higher than that of UGCA 276 and significantly lower than the mass of NGC 4214. Optical spectroscopy is available for the galaxy NGC 4214 that provides a possibility to compare its chemical characteristics to that of NGC 4163. Unfortunately, optical spectroscopy is not available for other neighbouring galaxy UGCA 276.

The stellar and gas masses of the neighboring dwarf galaxy NGC 4214 are $M_\star = 10^{8.9} M_\odot$ and $M_{HI} = 10^{8.5} M_\odot$ (Stilp et al., 2013). This results in a gas mass fraction of $\mu \sim 0.35$, and an oxygen abundance predicted by the simple cbm model of $12+\log(O/H)_{cbm} = 8.42$. Long-slit optical spectroscopy at 82 distinct spatial locations in NGC 4214 was carried out by Kobulnicky & Skillman (1996). They reveal no significant localized oxygen abundance differences and do not find large-scale variations in the oxygen abundance. The mean value of the oxygen abundance in NGC 4214 is $12 + \log(O/H) = 8.23$. The abundance distributions across the NGC 4214 was also considered in Pilyugin, Grebel, & Zinchenko (2015). There we found that this galaxy does not show a radial abundance gradient and that the mean value of the oxygen abundance is $12 + \log(O/H) = 8.2$. Comparison between the measured abundance and the one predicted by the simple cbm model for the chemical evolution of galaxies results in the conclusion that NGC 4214 kept around 0.6 of the potentially produced oxygen.

McQuinn et al. (2015) identified NGC 4163 as a post-starburst galaxy and NGC 4214 as a starburst galaxy. A fossil burst may imply a once significant galactic wind. On the other hand, McQuinn et al. (2010) infer a low absolute star formation rate during the past starburst in this galaxy, in contrast to, e.g., NGC 4214. Based on star formation histories derived from deep colour-magnitude diagrams obtained with the Hubble Space Telescope, McQuinn et al. (2010) estimate that only 2.8 ± 0.8 per cent of the stellar mass of NGC 4163 was created in its starburst a few hundred million years ago. They estimate the duration of the starburst to have lasted 460 ± 70 Myr, typical for many dwarf irregular galaxies in their sample and corresponding to about two dynamical time scales, which, as they discuss, argues against self-quenching. Combined with our findings, this may support a scenario of infall of little enriched gas as the main cause of the abundance variations. Analysis of the cosmological simulations show that the infall could lead to the removal of pre-existing gas, which is captured and mixed together with the infalling component (Khoperskov et al., 2021). Moreover, during infall gas can form extended ring-like structures. Such arc is clearly seen in the H$\alpha$ image (see Fig. 2 ) and HI image (Lelli et al., 2014) of NGC 4163.

## 5 | CONCLUSIONS

We analyzed publicly available IFU spectroscopic data of the dwarf irregular galaxy NGC 4163 obtained in the framework of the SDSS-IV MaNGA survey. We have measured the fluxes of the [O II]$\lambda 3727+\lambda 3729$, H$\beta$, [O III]$\lambda 4959$, [O III]$\lambda 5007$, [N II]$\lambda 6548$, H$\alpha$, [N II]$\lambda 6584$, [S II]$\lambda 6717$, and [S II]$\lambda 6731$ lines with a signal-to-noise ratio (SNR) higher than 3 for each line in 1176 spaxel spectra of NGC 4163 (and with SNR > 8 for 226 spaxel spectra).

Also, we detected auroral emission line [O II]$\lambda\lambda 7320,7330$ which is necessary for direct chemical abundance determination which allows us to measure chemical abundance by direct $T_e$ method.

A prominent feature of the NGC 4163 is that the $R_3$ line is significantly weaker in comparison to the spectra of other MaNGA galaxies and spectra of H II regions in nearby galaxies with a similar intensity of the $N_2$ line. As a result, the excitation of the spaxel spectra of the NGC 4163 is unusually low.

The mean value of the oxygen abundances in NGC 4163 obtained from the spaxel spectra through the $R$ calibration is 12 + log(O/H) = 7.55. This value is in agreement with the oxygen abundance obtained by Berg et al. (2012) through the direct $T_e$ method for one single region in this galaxy. On the other hand, our estimation of the oxygen abundance by direct method using our measurement of the O$^+$ auroral line 12+log(O/H) ∼ 7.8 is higher compared to the abundance found by Berg et al. (2012).

There is no evidence of the radial abundance gradient across NGC 4163, although the interstellar medium of this galaxy is not well mixed since the abundance scatter is large, ranging from 12 + log(O/H) ∼ 7.3 to ∼ 7.8.

An unusual negative trend between log(N/O) and oxygen abundance is detected. This trend is similar to the negative trend reported by Kumari et al. (2018) for NGC 4670 and some other dwarf galaxies.

The gas mass fraction in the NGC 4163 is low, $\mu \sim 0.25$. The low metallicity coupled with the low gas mass fraction implies that either the evolution of the galaxy NGC 4163 was accompanied by the strong galactic winds in the earlier epoch (since the oxygen abundance in the galaxy is only around 0.1 of the oxygen abundance potentially attainable in a galaxy of such astration level or gas mass fraction) or the abundance of the interstellar medium of the galaxy was reduced by pristine gas infall in the recent epoch. Further studies of circumgalactic medium is needed to distinguish between those two scenarios.

## ACKNOWLEDGMENTS

We are grateful to the referee for his/her constructive comments.

I.A.Z. and L.S.P. acknowledge support within the framework of Sonderforschungsbereich (SFB 881) on "The Milky Way System" (especially subproject A5), which is funded by the **German Research Foundation (DFG)**.

I.A.Z. and L.S.P. thank for hospitality of the Astronomisches Rechen-Institut at Heidelberg University, where part of this investigation was carried out.

L.S.P acknowledges support in the framework of the program "Support for the development of priority fields of scientific research" of the **NAS of Ukraine** ("ActivPhys", 2022-2023).

We acknowledge the usage of the HyperLeda database (http://leda.univ-lyon1.fr).

Funding for the Sloan Digital Sky Survey IV has been provided by the Alfred P. Sloan Foundation, the U.S. Department of Energy Office of Science, and the Participating Institutions. SDSS-IV acknowledges support and resources from the Center for High-Performance Computing at the University of Utah. The SDSS web site is www.sdss.org.

SDSS-IV is managed by the Astrophysical Research Consortium for the Participating Institutions of the SDSS Collaboration including the Brazilian Participation Group, the Carnegie Institution for Science, Carnegie Mellon University, the Chilean Participation Group, the French Participation Group, Harvard-Smithsonian Center for Astrophysics, Instituto de Astrofísica de Canarias, The Johns Hopkins University, Kavli Institute for the Physics and Mathematics of the Universe (IPMU) / University of Tokyo, the Korean Participation Group, Lawrence Berkeley National Laboratory, Leibniz Institut für Astrophysik Potsdam (AIP), Max-Planck-Institut für Astronomie (MPIA Heidelberg), Max-Planck-Institut für Astrophysik (MPA Garching), Max-Planck-Institut für Extraterrestrische Physik (MPE), National Astronomical Observatories of China, New Mexico State University, New York University, University of Notre Dame, Observatário Nacional / MCTI, The Ohio State University, Pennsylvania State University, Shanghai Astronomical Observatory, United Kingdom Participation Group, Universidad Nacional Autónoma de México, University of Arizona, University of Colorado Boulder, University of Oxford, University of Portsmouth, University of Utah, University of Virginia, University of Washington, University of Wisconsin, Vanderbilt University, and Yale University.